\documentstyle[aps,epsfig,multicol,amssymb,amsfonts]{revtex}
\tighten
\renewcommand{\widetext}
{\end{multicols}\global\columnwidth42.5pc}
\begin{document}
\newcommand{\be}{\begin{equation}}
\newcommand{\ee}{\end{equation}}
\newcommand{\bea}{\begin{eqnarray}}
\newcommand{\eea}{\end{eqnarray}}
\newcommand{\br}{{\bf r}}
\newcommand{\bk}{{\bf k}}
\newcommand{\bq}{{\bf q}}
\newcommand{\bn}{{\bf n}}
\draft
\title{Current correlations and quantum localization in 2D disordered
systems with broken time-reversal invariance}
\author{I.~V.~Gornyi$^{1,*}$, A.~D.~Mirlin$^{1,2,\dagger}$ and
P.~W\"olfle$^{1,2}$ }
\address{$^1$Institut
f\"ur Nanotechnologie, Forschungszentrum Karlsruhe, 76021 Karlsruhe,
Germany}
\address{$^2$Institut f\"ur Theorie der Kondensierten Materie,
Universit\"at Karlsruhe, 76128 Karlsruhe, Germany}
\date{\today}
\maketitle
\begin{abstract}
We study long-range correlations of equilibrium current
densities in a two-dimensional
mesoscopic system with the time reversal invariance
broken by a random or homogeneous magnetic field. Our result is
universal, i.e. it does not depend on the type (random potential or
random magnetic field) or correlation length of disorder. This
contradicts recent $\sigma$-model calculations of
Taras-Semchuk and Efetov (TS\&E) for the current correlation function,
as well as for the renormalization of the conductivity.
We show explicitly that the new term in the $\sigma$-model derived
by TS\&E  and claimed to lead to delocalization does not exist. 
The error in the derivation of TS\&E is traced to an incorrect
ultraviolet regularization procedure violating current conservation
and gauge invariance. 
\end{abstract}
\begin{multicols}{2}


The quantum coherence is well known to play a central role in the
physics of mesoscopic systems. It induces, in particular, long-range
spatial correlations of local densities of states and eigenfunction
amplitudes. The long-range character of these correlations is due to
the existence of massless modes, diffusons and Cooperons. This leads also
to strong mesoscopic fluctuations of global quantities, such as the
conductance or the inverse participation ratio.

In this paper, we present one more example of such long-range
correlations induced by the phase coherence of a system. Specifically,
we will calculate the correlation function of local equilibrium current
densities of electrons subject to a (possibly smooth) random potential (RP)
or random magnetic field (RMF). Non-zero local currents exist in a system if
the time reversal invariance is broken by a magnetic field, either
uniform or spatially fluctuating. We will demonstrate that the result
is independent of the type of disorder and has a rather universal
character.

This paper is to a large extent based on results obtained by two of us
(A.D.M. and P.W.) in 1994 in collaboration with the late Arkadii
Aronov. At 
that time our research was motivated by the paper \cite{zhang}, where
it was claimed that the above current fluctuations lead to
delocalization of electrons in random magnetic field, in contradiction
to our finding in \cite{amw}. Our results
remained however unpublished. In fact, the unfortunate error
in the calculation of Ref.~\cite{zhang} was quite obvious (these
authors confused the correlation function of equilibrium currents with
the current response function), and we did not see a necessity to
devote an additional publication to it. Also, evaluation of the
correlation function of local currents is quite straightforward and
follows essentially the same lines as the calculation of a typical
value of the global persistent current \cite{persist}.

However, in the meantime, the result of Ref.~\cite{zhang} has been
quoted in a number of publications as one of theoretical proposals
concerning the problem of localization 
of two-dimensional (2D) fermions in a
random magnetic field. Furthermore, the idea of \cite{zhang} was
recently revived, within a different line of reasoning, by
Taras-Semchuk and Efetov (TS\&E) \cite{taras}. Despite the criticism of
their work in the comment \cite{comment}, TS\&E have made public a
rather extended paper \cite{taras1}
presenting in detail their calculations and
conclusions. If true, their results, predicting a delocalizing one-loop
contribution to the $\beta$-function governing the scaling of
conductance for systems with broken time
reversal invariance (unitary symmetry class), would change completely
our understanding of localization properties of these systems. In
particular, they would force one to reconsider the phase diagram of
the quantum Hall effect. Most noteworthy, the findings of TS\&E
challenge the notion of universality for this class of systems: the new
contribution to the $\beta$-function derived by them has a
non-universal coefficient $\beta_{\rm TSE}$ (see below) depending on
the type (RP or RMF) and the correlation length of the disorder. 
We thus believe that it is timely to present
our results on current density correlations and our criticism of
Refs.~\cite{taras,taras1} without the length restrictions implied by the
Comment format.


We consider the correlation function of local current
densities $j_\alpha(\br,E)$ at equilibrium in 2D (we set $\hbar=1$ and
omit the spin degree of freedom),
\bea
I_{\alpha \beta}(\br&-&\br^\prime,\omega)\equiv
\langle j_\alpha(\br,E+\omega) j_\beta(\br^\prime,E) \rangle
\label{defI} \\
{\bf j}(\br,E)&=&
\frac{e}{2 m}[i\nabla_{\br}-i\nabla_{\tilde \br}+{2e\over c} {\bf A}(\br)]
\nonumber \\
&\times&\left.\frac{1}{2\pi i} 
\left[G^R_E(\br,{\tilde \br})- G^A_E(\br,{\tilde \br})\right] 
\right|_{{\tilde \br}=\br}\ ,
\eea
where $G^R_E,G^A_E$ are retarded and advanced Green's functions,
and $\langle \ldots \rangle$ denotes disorder (RP or RMF) averaging.
In view of the current conservation, $\nabla{\bf j} = 0$, the function
$I_{\alpha \beta}$ is transverse in momentum space,
\be
I_{\alpha \beta}(\bq,\omega)=
e^2 F(q,\omega)
\left(\delta_{\alpha \beta} - \frac{q_\alpha q_\beta}{q^2} \right).
\label{transv}
\ee
Since we are interested in long-range current correlations,
we will study the low-momentum, low-frequency behavior of $I_{\alpha
\beta}$. A finite value of $F(q,\omega)$ in the low-$q$ limit would
imply dipole-type long-range correlations in coordinate space,
\be
I_{\alpha \beta}(\bbox{\rho}) \propto
\nabla_\alpha \nabla_\beta \ln\rho=
\frac{1}{\rho^2}
\left(\delta_{\alpha \beta} -
2\frac{\rho_\alpha \rho_\beta}{\rho^2} \right).
\label{Iincoord}
\ee
We will make use of the representation of the correlation function
$I_{\alpha \beta}$ in terms of a functional derivative of a generating
functional ${\cal F}\{{\bf A},{\bf A^\prime}\} $ with respect to
vector-potential source fields ${\bf A}(\br)$, ${\bf A}^\prime(\br)$,
\be
I_{\alpha \beta}(\br-\br^\prime)= \left.
c^2\frac{\delta}{\delta A_{\alpha}(\br)}
\frac{\delta}{\delta A^\prime_{\beta}(\br^\prime)}
{\cal F}\{{\bf A},{\bf A^\prime}\}\right|_{{\bf A}={\bf A^\prime}=0},
\label{IderivA}
\ee
where
\bea
{\cal F}\{{\bf A},{\bf A^\prime}\}&=&
\left\langle\left[{\rm Tr} \ln G^R_E\{{\bf A}\}-
{\rm Tr} \ln G^A_E\{{\bf A}\} \right]\right.
\nonumber \\
&\times& \left. \left[{\rm Tr} \ln G^R_{E+\omega}\{{\bf A^\prime}\}-
{\rm Tr} \ln G^A_{E+\omega}\{{\bf A^\prime}\} \right] \right \rangle.
\label{FAA}
\eea
It is seen from Eq.~(\ref{IderivA}) that the transverse character
of $I_{\alpha \beta}$ is intimately related to gauge invariance of the
generating functional ${\cal F}\{{\bf A},{\bf A^\prime}\}.$

We start from a diagrammatic analysis in the framework of a conventional
impurity diagram technique. The diagrams for ${\cal F}$ are obtained by
connecting two closed electron loops by impurity lines;
the diagrams for $I_{\alpha \beta}$ are then generated by
inserting two current vertices (one in each loop)
in all possible ways.

Let us first consider the case of a random scalar potential
with a correlation function
$\langle U(\br) U(\br^\prime) \rangle =W(\br-\br^\prime)$;
its correlation length $d$ is assumed to be finite, though it
can be arbitrary as compared to the wave length $\lambda_{\rm F}$.
In particular, if $d \gg \lambda_{\rm F}$ (smooth RP), the scattering is
of a small-angle nature, implying that the transport relaxation time
is much longer than the single-particle one, $\tau_{\rm tr} \gg \tau$.
It is clear that  in any finite order $n$ of the perturbative
expansion ($n$ is a number of impurity lines connecting the
fermion loops), the current correlation function
$I_{\alpha \beta}^{(n)}$ is finite-ranged, i.e., it has a finite
correlation length, beyond which the correlations decay exponentially
in view of the finite-range character
of $W(\br-\br^\prime)$.
This means that in momentum space $I_{\alpha \beta}^{(n)}(\bq)$
should have no singularity as $q \to 0$, implying that
$F(q)\propto q^2$. Indeed, the sum of all the diagrams of the 
$n$-th order can be presented in the form
\bea
I_{\alpha \beta}^{(n)}(\bq) &=&
\frac{1}{n!}\int (d\bq_1)\ldots (d\bq_{n-1})  W(q_1)\ldots W(q_n)
\nonumber \\
& \times &
 T_{\alpha}(\bq_1,\ldots,\bq_n)
T_\beta(-\bq_1,\ldots,-\bq_n),
\label{RPpt}
\eea
where $(d\bq)=d^2q/(2\pi)^2$, and 
$T_{\alpha}(\bq_1,\ldots,\bq_n)$ denotes a vertex part to which $n$
impurity lines with momenta $\bq_1,\ldots,\bq_n$ satisfying
$\sum_i\bq_i=\bq$ are attached in all possible ways.
Because of its vector character and the symmetry with respect to permutation
of $\bq_i$, the block $T_{\alpha}$ should be proportional to $q$;
the current conservation implies its transverse character
$T_{\alpha}\propto {\bar q}_\alpha$, where
${\bar q}_\alpha = \epsilon_{\alpha \beta} q_\beta$
($\epsilon_{\alpha \beta}$ is the antisymmetric tensor).
When substituted in Eq.~(\ref{RPpt}), this yields again
\be
F_{\rm {\small RP}}(q)\propto q^2\ , \qquad q\to 0.
\label{FRPq2}
\ee

We now turn to the RMF case. The situation now is
somewhat less trivial, since although the RMF
correlation function
$W_B(\br-\br^\prime)=\langle B(\br) B(\br^\prime)\rangle$
is assumed to be of a finite range, the correlation function of the
corresponding vector potential $A(\br)$ is long-ranged,
which is reflected by a singularity at $q=0$ in momentum space,
\be
{\cal W}_{\alpha\beta}({\bf q})=\frac{W_B(q)}{q^2}
\left(\delta_{\alpha\beta}-\frac{q_\alpha q_\beta}{q^2}\right)\ .
\label{corrAA}
\ee
Let us, however, analyze the RMF analog of Eq.~(\ref{RPpt}),
\bea
\label{RMFpt}
I_{\alpha \beta}^{(n)}(\bq)&=&\frac{1}{n!}\int (d\bq_1)\ldots
(d\bq_{n-1}) \nonumber \\
&\times & {\cal W}_{\alpha_1\beta_1}(\bq_1)\ldots
{\cal W}_{\alpha_n\beta_n}(\bq_n) \nonumber \\
&\times& T_{\alpha \alpha_1 \ldots \alpha_n}(\bq_1,\ldots,\bq_n)
\nonumber\\ &\times&
T_{\beta \beta_1 \ldots \beta_n}(-\bq_1,\ldots,-\bq_n)\ . 
\eea
As in the RP case, the block T represents a sum of all possible
diagrams with one current vertex and $n$ disorder lines attached.
This includes, in addition to the diagrams appearing in the RP case,
diagrams having vertices originating from the ${\bf A}^2$ term in the 
Hamiltonian, ${\hat H}=(-i\nabla-e{\bf A}(\br)/c)^2/2m$ 
(i.e. with two RMF lines joining the electron line at the same point,
or with a RMF line starting at the external current vertex). Such 
diagrams, while being of minor importance for the diffusion
contribution considered below (they are suppressed by the factor 
$(k_{\rm F} l_{\rm tr})^{-1},$ where $k_{\rm F}$ is the Fermi momentum
and $l_{\rm tr}=v_{\rm F}\tau_{\rm tr}$),
are crucially important in the perturbation theory \cite{footnote2}. 
It is not difficult to see that all the diagrams for the 
vertex part $T$ in Eq.~(\ref{RMFpt}) can be generated by
variation of a current
with respect to the vector potential
\bea
&& T_{\alpha \alpha_1 \ldots \alpha_n}(\bq_1,\ldots,\bq_n)
\nonumber \\ && \ \ \left. 
=\frac{\delta}{\delta {\tilde A}_{\alpha_1}(\bq_1)}\ldots
\frac{\delta}{\delta {\tilde A}_{\alpha_n}(\bq_n)}
\langle j_\alpha ({\bf A}+{\bf {\tilde A}}) \rangle
\right|_{{\tilde A}=0}.
\label{ddA}
\eea
In view of the gauge invariance, the vertex part in the limit $q\to 0$
must have the form
\be
T_{\alpha\alpha_1\ldots\alpha_n}({\bf q}_1,\ldots,{\bf q}_n)\propto
{\bar q}_\alpha ({\bar q_1})_{\alpha_1}\ldots ({\bar q_n})_{\alpha_n}\ ,
\label{Trmf}
\ee
so that all the singularities in the vector potential correlators
${\cal W}_{\alpha_i\beta_i}(q_i)$ are canceled by the
vertex parts, yielding
\be
F_{\rm {\small RMF}}(q)\propto q^2\ , \qquad q \to 0\ .
\label{FRMFq2}
\ee
Therefore, in any order $n$ the perturbative contribution to the current
correlator is of finite range in the RMF case as well, despite the
singular nature of the vector potential correlator (\ref{corrAA}).

\begin{figure}
\narrowtext
\centerline{ {\epsfxsize=8cm{\epsfbox{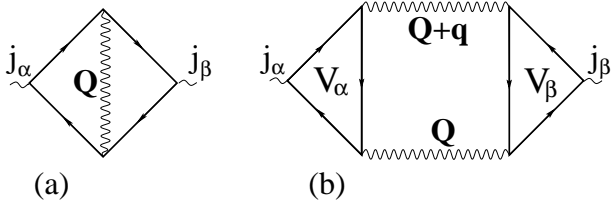}} }}
\vspace{3mm}
\caption{  Leading diffuson diagrams for the current correlator
 with one  (a) and two (b) diffusons. }
\label{fig1}
\end{figure}

\begin{figure}
\narrowtext
\centerline{ {\epsfxsize=5cm{\epsfbox{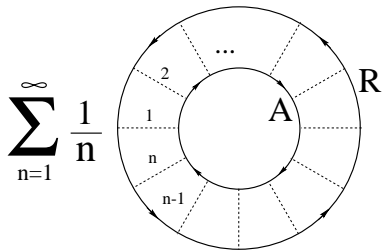}} }}
\vspace{3mm}
\caption{Generating functional $\cal F$ for the diagrams of
Fig.~\ref{fig1}.  } 
\label{fig2}
\end{figure}

As usual in mesoscopic physics, long-range correlations are determined
by a diffuson contribution. We assume Cooperons to be suppressed by
the random or homogeneous magnetic field; in the latter case the
magnetic field is assumed to be classically weak ($\omega_c\tau_{\rm
tr}\ll 1$). The corresponding diagrams are
shown in Fig.~\ref{fig1}. The sum of these two diagrams satisfies the current
conservation requirement, since they can be obtained from
Eq.~(\ref{IderivA}) with the generating functional ${\cal F}$ shown in
Fig.~\ref{fig2}.


In order to take into account anisotropic scattering we use the
following expression for the diffusons \cite{wb}
\be
D_{\bk \bk^\prime}(\bq,\omega)=
{1\over 2\pi\nu\tau^2}
\frac{\gamma_{\bk}(\bq)\gamma_{\bk^\prime}(\bq)}{Dq^2-i\omega}
+{\cal D}_{\rm reg},
\label{diffuson}
\ee
where $\nu=m/2\pi$ is the density of states, 
$D=v_{\rm F}^2 \tau_{\rm tr}/2$ is the diffusion constant,
$\gamma_{\bk}(\bq)=1-i(\tau_{\rm tr}-\tau)\bk\bq/m,$
and
$$D_{\rm reg}=\sum_{m\neq 0}\frac{\lambda_m}{1-\lambda_m}
\chi_m(\bk)\chi^*_m(\bk^\prime),$$
with $\chi_m(\bk)$, $\lambda_m$ being the eigenfunctions and eigenvalues of
the scattering operator with the kernel
$W(\bk,\bk^\prime)|G^R(k^\prime)|^2$.
When calculating the contribution of the diagram Fig.~\ref{fig1}a,
only the 
singular part of the diffuson (the first term in (\ref{diffuson}) )
should be taken into account, yielding
\bea
I^a_{\alpha \beta} & = & \frac{e^2}{2\pi^2}{\rm Re}
\int (d\bk)(d{\bf Q}) v_\alpha v_\beta
[G^R(k)G^A(k)]^2 D_{\bk \bk}(Q,\omega) \nonumber \\
&=& \delta_{\alpha \beta}\frac{e^2}{2\pi^2}
v_{\rm F}^2\tau {\rm Re}\int_{0}^{l_{\rm tr}^{-1}}
\frac{(d{\bf Q})}{D{\bf Q}^2-i\omega} \nonumber \\
&=& \delta_{\alpha \beta}\frac{e^2}{2\pi^3}\frac{\tau}{\tau_{\rm tr}} 
\ln \frac{L_{\omega}}{l_{\rm tr}}\ ,
\label{fig2a}
\eea
where $L_{\omega}=(D/\omega)^{1/2}$. Let us note that we consider an
infinite system; for a finite sample 
of size $L<L_\omega$ one should replace $L_{\omega}$ by the system
size $L$, which plays then the role of the infrared (IR) cut-off.

The diagram Fig.~\ref{fig1}b gives
\bea
I^b_{\alpha \beta}&=&\frac{e^2}{2\pi^2}{\rm Re}
\int (d\bk) (d\bk^\prime) (d{\bf Q}) \nonumber \\
& \times & V_\alpha(\bk,{\bf Q},\bq)
[V_\beta(\bk^\prime,-{\bf Q},-\bq)]^* \nonumber \\
&\times & D_{\bk \bk^\prime}({\bf Q},\omega)
D_{\bk \bk^\prime}({\bf Q} + \bq,\omega),
\label{fig2b}
\eea
where $V_{\alpha(\beta)}$ are the triangle vertex parts,
\be
V_\alpha(\bk,{\bf Q},\bq)=v_\alpha
G^R(\bk+\bq)G^A(\bk-{\bf Q})G^R(\bk).
\label{Sparts}
\ee
Substituting (\ref{diffuson}) in (\ref{fig2b}), we find the following
two relevant contributions.
If one takes into account the singular parts of both diffusons,
then the Green functions can be expanded in small
momenta $\bq$ and ${\bf Q}$, which yields for $q\gg L_\omega$
\bea
I^{b1}_{\alpha \beta}&=&-\frac{e^2 D^2}{2\pi^2}
{\rm Re}\int_{0}^{l_{\rm tr}^{-1}}\!
\frac{(d{\bf Q})(2Q_\alpha+q_\alpha)(2Q_\beta+q_\beta)}
{[D{\bf Q}^2-i\omega][D({\bf Q}+\bq)^2-i\omega]}
\nonumber \\
&=&\frac{e^2}{2\pi^3}
\left[\delta_{\alpha \beta}\ln(ql_{\rm tr}) -
\frac{q_\alpha q_\beta}{q^2}\ln(q L_\omega) \right].
\label{Ib1}
\eea
Another contribution
\be
I^{b2}_{\alpha \beta}
=\delta_{\alpha \beta}\frac{e^2}{2\pi^3}\frac{\tau_{\rm tr}-\tau}
{\tau_{\rm tr}}
\ln \frac{L_\omega}{l_{\rm tr}}
\label{Ib2}
\ee
arises when the regular part of one of the two diffusons
is taken into account.

Combining Eqs.~(\ref{fig2a}), (\ref{Ib1}) and (\ref{Ib2}),
we finally get
\be
I_{\alpha \beta}=\frac{e^2}{2\pi^3}\ln(q L_\omega)
\left(\delta_{\alpha \beta} - \frac{q_\alpha q_\beta}{q^2} \right),
\label{diffres}
\ee
which has the required transverse form (\ref{transv})
with
\be
F(q,\omega)=\frac{1}{2\pi^3} \ln(q L_\omega)\ , \qquad
L_\omega^{-1} \ll q \ll l_{\rm tr}^{-1}\ .
\label{Fdiff}
\ee
Note, that neither the  single-particle
nor the transport scattering time enter the final result, which is
universal and does not depend on the character of the disorder.

Our result (\ref{diffres}) differs from that of  TS\&E (Ref.~\cite{taras1})
in two crucial aspects. First of all, in the formula by TS\&E
the factor $F(q,\omega)\sim\ln(q L_\omega)$ is replaced by a large
constant $\beta_{\rm TSE} \gg 1$. Specifically, 
for the RMF case  TS\&E find
\be
\beta_{\rm TSE}\sim\frac{\tau_{\rm tr}}{\tau},
\label{betaRMF}
\ee
while for a smooth RP their consideration yields \cite{footnote1}
\be
\beta_{\rm TSE}\sim\left(\frac{\tau_{\rm tr}}{\tau}\right)^{1/2}.
\label{betaRP}
\ee
Secondly, while our contribution comes from the IR region
$Q\ll\l_{\rm tr}^{-1}$, their term is claimed to be determined by the
ultraviolet (UV) region $Q \gg \l_{\rm tr}^{-1}$. We will now show that
the result of  TS\&E is wrong.

In fact, already the above considerations imply that the findings of
TS\&E are incorrect. Indeed, we have demonstrated that
$F(q,\omega)\propto q^2$ as $q \to 0$ \cite{footnote}.
In other words, Eq.~(\ref{IderivA}) implies $F(q=0,\omega)=0$,
since a gauge-invariant generating functional ${\cal F}$ cannot depend
on a change of the vector potential by a constant. Therefore the
TS\&E result $F(q=0,\omega)=\beta_{\rm TSE}\gg 1$ cannot be correct.

To make closer contact with the work of TS\&E and, in particular,
to show explicitly their mistake, we will now rederive Eq. (\ref{diffres})
in the framework of the ballistic $\sigma$-model
approach \cite{MK,AASA}. The $\sigma$-model can be derived by
averaging over a smooth RP or a RMF following \cite{wb,amw}, as
explained in detail in \cite{taras1}. 
Performing a standard calculation and treating the
 $\sigma$-model correlator perturbatively, one gets the leading
contribution in the form
\bea
&& I_{\alpha \beta}(\br-\br^\prime,\omega) = \frac{e^2}{2\pi^2}v_{\rm F}^2
\int \frac{d{\bf n} d{\bf n}^\prime}{(2\pi)^2} \nonumber \\ 
&& \qquad \times  
n_\alpha g(\br,{\bf n}; \br^\prime, {\bf n}^\prime;\omega)
n_\beta g(\br^\prime,{\bf n}^\prime; \br, {\bf n};\omega),
\label{IGG}
\eea
where $g(\br,{\bf n}; \br^\prime, {\bf n}^\prime;\omega)$ is
a propagator describing the motion from the point 
$(\br^\prime,{\bf n}^\prime)$  of the
phase space to the point  $(\br,{\bf n})$, with the unit vector 
${\bf n}$ staying for the velocity direction of a particle on the
Fermi surface. The propagator $g$ obeys the Liouville--Boltzmann
equation 
\bea
&& ({\cal L}-i\omega)
g(\br,{\bf n};\br^\prime,{\bf n}^\prime;\omega) \nonumber\\
&& - \int  \frac{d{\bf n}^{\prime\prime}}{2\pi} 
w({\bf n},{\bf n}^{\prime \prime}) 
[g(\br,{\bf n}^{\prime \prime};\br^\prime,{\bf n}^\prime;\omega)
 -g(\br,{\bf n};\br^\prime,{\bf n}^\prime;\omega)] 
 \nonumber \\
&& = \delta(\br-\br^\prime)\delta({\bf n}-{\bf n}^\prime),
\label{LG}
\eea
where ${\cal L}=v_{\rm F} {\bf n} \nabla_{\br}$ and
$w({\bf n},{\bf n}^{\prime})$ is the scattering cross-section ($\phi$
is the angle between ${\bf n}$ and ${\bf n'}$),
\be
w({\bf n},{\bf n}^{\prime})=\left\{  \begin{array}{ll}
\! 2\pi\nu W(2p_{\rm F}\sin{\phi\over 2}), & \ \ {\rm RP}, \\
\! 2\pi\nu \cot^2{\phi\over 2}\left({e\over 2mc}\right)^2  
W_B(2p_{\rm F}\sin{\phi\over 2}), & \ \ {\rm RMF}.
\end{array} \right.
\label{crossec}
\ee

Note that (\ref{IGG}) is identical to the Altshuler--Shklovskii--type
contribution to the local density of states correlation function
\cite{AASA}, up to the velocity factors $v_{\rm F} n_\alpha,
v_{\rm F} n_\beta$. 

Equation (\ref{IGG}) has the same form as the formula (37)
of TS\&E, Ref. \cite{taras1}.
In fact, there is a slight difference between the two
formulas: while our function $g$ is a full propagator of
the Liouville--Boltzmann equation, the function
$\Gamma(\br,{\bf n}; \br^\prime, {\bf n}^\prime)$
of TS\&E is obtained by projecting this equation onto
the space of non-zero harmonics in the velocity space.
As a consequence, our propagator has a diffusive behavior
at large distances (see below), while $\Gamma$ of TS\&E
does not. This is because we treat the $\sigma$-model
field fully perturbatively, while TS\&E do this with respect to
the non-zero harmonics only. Therefore some of the contributions
to our Eq.~(\ref{IGG}) are shifted by them into other terms not included
in Eq.~(37) of Ref. \cite{taras1}.
While being crucial for a proper account of the diffusive contribution,
this difference is irrelevant for the analysis of the contribution
of the ballistic region, with characteristic momenta of the
propagators satisfying $Q\gg l_{\rm tr}^{-1}$, which is claimed by TS\&E
to be responsible for the results of Ref. \cite{taras,taras1}.

In the hydrodynamic limit ($Ql_{\rm tr}\ll 1$,
$\omega\tau_{\rm tr}\ll 1$) the propagator
$g({\bf Q};{\bf n},{\bf n}^\prime;\omega)$
has the conventional diffusive behavior (in view
of the vector nature of the factors $n_\alpha, n_\beta$
in (\ref{IGG}) we have to take into account the leading
correction in $Ql_{\rm tr}$),
\be
g({\bf Q};{\bf n},{\bf n}^\prime;\omega)=
\frac{(1-il_{\rm tr}{\bf Q}{\bf n})
      (1-il_{\rm tr}{\bf Q}{\bf n}^\prime)}{DQ^2-i\omega}\ .
\label{gdiff}
\ee
Substituting (\ref{gdiff}) in (\ref{IGG}), we arrive again at
Eq.~(\ref{diffres}).

\begin{figure}
\narrowtext
\centerline{ {\epsfxsize=5cm{\epsfbox{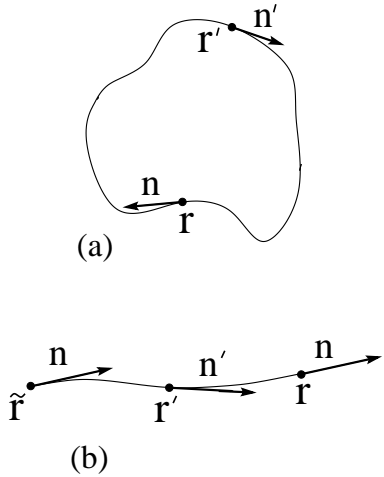}} }}
\vspace{3mm}
\caption{(a) The paths contributing to the current correlation
function $I_{\alpha\beta}$ , Eq.~(\ref{IGG}); (b) paths giving rise to
the spurious 
contribution to $I_{\alpha\beta}$ found by TS\&E.  } 
\label{fig3}
\end{figure}

Let us now demonstrate how the statement $F(q=0,\omega)=0$
is obtained within this approach. This can be most clearly done if
one introduces the path integral representation for the
propagator
\bea
g(\br,{\bf n}; \br^\prime, {\bf n}^\prime; \omega)&=&
\int^{\infty}_0 dT e^{i \omega T}
g(\br,{\bf n}; \br^\prime, {\bf n}^\prime; T); \nonumber \\
g(\br,{\bf n}; \br^\prime, {\bf n}^\prime;T)&=&
\int^{{\bf z}(T)=\{\br,{\bf n}\}}_{{\bf z}(0)=\{\br^\prime,{\bf n}^\prime\}}
{\cal D}{\br}(t)e^{iS[\br(t)]},
\label{path}
\eea
where ${\dot{\br}}(t)=v_{\rm F} \bn(t)$ and ${\bf z}(t)=\{\br(t),\bn(t) \}$.
Here $S$ is the action corresponding to the dynamics determined by
Eq.~(\ref{LG}). 
Using (\ref{path}), it is easy to show that
\bea
&& \int d{\br}^\prime g(\br,{\bf n}; \br^\prime, {\bf n}^\prime; \omega)
n_\beta^\prime
g(\br^\prime,{\bf n}^\prime;{\tilde\br},{\tilde\bn};\omega)
\nonumber \\
&& =\int_0^\infty dT e^{i\omega T}
\int^{{\bf z}^\prime(T)=\{\br,{\bf n}\}}
_{{\bf z}^\prime(0)=\{{\tilde\br},{\tilde\bn}\}}
{\cal D}{\br^\prime}(t)
\int_0^T dt\, n^\prime_\beta(t) e^{iS[\br^\prime(t)]}\nonumber \\
&& =g(\br,{\bf n};{\tilde\br},{\tilde\bn};\omega)
\frac{(\br - {\tilde\br} )_\beta}{v_{\rm F}}\ .
\label{svpath}
\eea
Substituting this (with $\br = {\tilde\br}$) in
Eq.~(\ref{IGG}) integrated over $\br'$, we immediately get
$I_{\alpha \beta}(q=0,\omega)=0$, in agreement with our earlier
analysis. This can be traced back to the fact that an integral of
velocity vector over a closed loop is equal to zero, see
Fig.~\ref{fig3}a. Note that one should exclude from Eq.~(\ref{svpath})
a contribution of a straight trajectory of zero length in the limit
$\tilde{\bf r}\to{\bf r}$ (see a detailed discussion below). 
In fact, this proof of $F(q=0,\omega)=0$ can be cast in the
form fully analogous to the above diagrammatic derivation
based on Eq.~(\ref{IderivA}). Indeed, Eq. (\ref{IGG}) can be
rewritten in the form (\ref{IderivA}) with the generating
functional given by an integral over closed paths in the phase space,
\bea
&& {\cal F}\left\{ {\bf A},{\bf A}^\prime \right \}=
\int_0^\infty {dT\over T} 
e^{i\omega T} \int_{{\bf z}(0)={\bf z}(T)} {\cal D}\br(t)
\nonumber \\
&&\times \exp\left(iS[\br(t)]-i{e\over c} \int dt\, {\dot \br}(t)
\left[{\bf A}(\br(t))+{\bf A}^\prime(\br(t))\right]\right). 
\nonumber\\ &&
\label{FAApath}
\eea
The trajectories $\br(t)$ in Eqs.~(\ref{svpath}), (\ref{FAApath})
are classical counterparts of the diffuson ladders in
Fig.~\ref{fig1}.  
Since the integral in (\ref{FAApath}) goes over closed loops, the
functional ${\cal F}$ is gauge-invariant (the second term in the
action yields the flux through the loop),
which immediately leads to $F(q=0,\omega)=0$ as explained above.


Why do TS\&E fail to observe that $F(q=0,\omega)$ is identically
zero and find instead the results (\ref{betaRMF}), (\ref{betaRP})
for this quantity? The reason is as follows. TS\&E evaluate Eq.~(\ref{IGG})
by writing it in momentum space and imposing an UV cut-off
$Q_0\gg l_{\rm tr}^{-1}$ which is set $Q_0=(v_{\rm F}\tau)^{-1}$
in the end of the calculation. It is easy to see that within this procedure
Eq. (\ref{IGG}) is replaced by
\be
I^{\rm TSE}_{\alpha \beta}(\br-\br^\prime, \omega)=
\frac{e^2v^2_{\rm F}}{2\pi^2} \int d{\tilde\br}
\Delta(\br-{\tilde\br},Q_0)
{\cal J}_{\alpha \beta}(\br,\br^\prime,{\tilde\br};\omega),
\label{ITSE}
\ee
where
\bea
&& {\cal J}_{\alpha \beta}(\br,\br^\prime,{\tilde\br};\omega)
= \int \frac{d{\bf n}d{\bf n}^\prime}{(2\pi)^2} \nonumber \\
&& \qquad\times  n_\alpha
g(\br,{\bf n}; \br^\prime, {\bf n}^\prime;\omega)
n^\prime_\beta
g(\br^\prime,{\bf n}^\prime; {\tilde \br}, {\bf n}; \omega)\ ,
\label{calJ}
\eea
and  $\Delta(\bbox{\rho},Q_0)$ is a broadened
$\delta$-function,
\be
\Delta(\bbox{\rho},Q_0)=\int_{|{\bf Q}|<Q_0} 
(d{\bf Q})e^{i\bbox{Q\rho}}=
\frac{Q_0 J_1(Q_0 \rho)}{2\pi \rho}\ ,
\label{deltaQ0}
\ee
with $J_1(x)$ the Bessel function. 
The crucial difference between Eq.~(\ref{ITSE}) and the exact
formula (\ref{IGG}) is that the former is determined by {\it non-closed}
paths with $|{\tilde\br}-\br|\sim Q_0^{-1}$.
In particular, there is a contribution to (\ref{ITSE}) from
a very short, almost straight path of a length
$\simeq|{\tilde\br}-\br|$,
with the point $\br^\prime$ lying between the points
$\br$ and ${\tilde\br}$ (see Fig.~\ref{fig3}b).
It is precisely this spurious contribution that
determines the result of TS\&E. To demonstrate this, we simply calculate
this contribution. According to (\ref{ITSE}), evaluation of the $q\to
0$ limit of $I_{\alpha\beta}^{\rm TSE}(q,\omega)$ involves integration of 
${\cal J}_{\alpha\beta}$ over $\br'$.  
Using (\ref{svpath}), we easily find that the contribution of the
almost-direct trajectories to $\int d\br^\prime{\cal J}_{\alpha\beta}$
is equal to
\be
\int d\br^\prime
{\cal J}_{\alpha\beta}(\br,\br^\prime,{\tilde\br};\omega)=
\int\frac{d\bn}{2\pi}
g(\br,{\bf n};{\tilde\br},{\bf n};\omega)
\frac{\rho_\alpha \rho_\beta}{v_{\rm F} \rho},
\label{direct}
\ee
where ${\bbox \rho} = \br-{\tilde\br}$.
If one completely neglects the disorder, $g$ becomes the free
propagator (since we discuss a ballistic contribution,
the frequency $\omega$ is fully irrelevant assuming
$\omega < \tau_{\rm tr}^{-1}$, and one can  simply set $\omega=0$),
\be
g_0(\br,{\bf n}; \br^\prime, {\bf n}^\prime)=
\frac{1}{2\pi v_{\rm F}|\br-\br^\prime|}
\delta\left({\bf n}-\frac{\br-\br^\prime}{|\br-\br^\prime|}\right)
\delta({\bf n}-{\bf n}^\prime),
\label{clprop}
\ee
substitution of which into (\ref{direct}) leads to a divergency
of the type $\delta(0)$. To see how this divergency is smeared
by the disorder scattering, we rewrite (\ref{direct}) in the
form
\be
\int d\br^\prime
{\cal J}_{\alpha\beta}(\br,\br^\prime,{\tilde\br};\omega)=
\frac{\rho_\alpha \rho_\beta}{v_{\rm F}^2\rho^2}
{\cal P}(0, \rho/v_{\rm F}),
\label{smeared}
\ee
where ${\cal P}(\delta\phi,t)$ is the distribution of
the variation $\delta\phi$ of the velocity angle in time $t$.
In the case of a RP the velocity angle diffuses with the diffusion
constant $\tau_{\rm tr}^{-1}$,
\be
{\cal P}_{RP}(\delta\phi,t)=
\left({\tau_{\rm tr}\over 4 \pi t}\right)^{1/2}\exp\left[
-{\tau_{\rm tr}\over 4t}(\delta\phi)^2\right]\ ,
\label{PRP}
\ee
while in a RMF one has a Cauchy distribution \cite{AMW}
\be
{\cal P}_{RMF}(\delta\phi,t)=\frac{1}{\pi}
\frac{2t/\tau_{\rm tr}}{(2t/\tau_{\rm tr})^2+(\delta\phi)^2}.
\label{PRMF}
\ee
Substituting (\ref{PRP}), (\ref{PRMF}) in (\ref{smeared})
we arrive at
\bea
\int d\br^\prime
&& {\cal J}_{\alpha\beta}(\br,\br^\prime,{\tilde\br};\omega)\nonumber\\
&& \qquad  =
\frac{1}{v_{\rm F}^2} \frac{\rho_\alpha
\rho_\beta}{\rho^2} 
\times \left\{
\begin{array}{ll}
(l_{\rm tr}/4\pi\rho)^{1/2}, & \ {\rm RP} \\
l_{\rm tr}/2\pi\rho, & \ {\rm RMF}.
\end{array}
\right.
\label{calJrho}
\eea
Finally, substituting (\ref{calJrho}) in (\ref{ITSE}),
we get
\bea
I^{\rm TSE}_{\alpha \beta}(q=0,\omega) &=&
\delta_{\alpha\beta}\frac{e^2}{2\pi^3}\nonumber \\
&\times& \left\{
\begin{array}{ll}
\gamma(Q_0 l_{\rm tr})^{1/2}, & \ {\rm RP} \\
Q_0 l_{\rm tr}/4, & \ {\rm RMF},
\end{array}
\right.
\label{ITSEq}
\eea
with a numerical coefficient $\gamma=\sqrt{\pi/2}
[\Gamma({3\over 4})/4\Gamma({5\over 4})]$,
reproducing the results (\ref{betaRMF}), (\ref{betaRP}) of TS\&E.

We have thus demonstrated that the findings of TS\&E result from
an incorrect regularization procedure. Note that their term (\ref{ITSEq})
is not transverse with respect to $q$ and thus violates the current
conservation. One could anticipate this without any calculations:
a non-zero $q\to 0$ limit of $I_{\alpha\beta}$ in combination with
current conservation would imply, according to (\ref{transv}),
existence of a singular term $\sim q_\alpha q_\beta/q^2$
of a long-range nature (\ref{Iincoord}) in coordinate space.
However, TS\&E deal with short scales $\ll l_{\rm tr}$
only and thus have no chance to obtain such a term.
This violation of current conservation remains, however,
hidden in \cite{taras1}, where only the trace
$I^{\rm TSE}_{\alpha\alpha}(q=0,\omega=0)$ is calculated,
while the transverse structure of $I^{\rm TSE}_{\alpha\beta}$
is then ``restored'' by hand.

The violation of current conservation is a direct consequence
of the violation of gauge invariance of the generating
functional ${\cal F}$ by the regularization of TS\&E. Indeed,
this gauge invariance is guaranteed by the fact that only
closed paths contribute to (\ref{FAApath}) while TS\&E allow for
non-closed short paths which produce their spurious term.
It is clear from our discussion what would be a proper UV
regularization procedure respecting the gauge invariance
constraint. One should discard the contribution of those 
closed loops to the effective action which have a length shorter than
a cut-off length $v_FT_0$, which amounts to setting the lower limit of
the $T$--integral in (\ref{FAApath}) equal to $T_0$. Note that for a
smooth 
disorder ($d\gg\lambda_F$) the quasiclassical result (\ref{IGG}) for
the current correlator will be essentially independent of the cut-off
$T_0\ll\tau_{\rm tr}$, since a particle cannot make a closed loop in a
time much shorter than $\tau_{\rm tr}$ (if one neglects exponentially
rare events). 
 

To shed more light on the cause for the appearance of the
contribution $I_{\alpha\beta}^{\rm TSE}$, let us note that the length
of the relevant 
trajectories $\rho \sim 1/Q_0$ goes to zero when the cut-off is removed,
$Q_0 \to \infty$. Such ``zero-length trajectories'' are well known to 
appear in the quasiclassical representation of the Green's function 
$G_E(\br,{\tilde \br})$ at ${\tilde \br} \to \br$. 
It is also well known that the 
contribution of such a direct trajectory of length zero 
should be split off
from the remaining sum over periodic orbits in the trace formula
for  $G_E(\br,\br)$; the former yields the average (Weyl)
contribution to the density of states, while the latter describes 
fluctuations around it \cite{gut}. In the present context of current 
correlations, the contribution of the zero-length trajectory is zero
since $\langle {\bf j}(\br, E) \rangle = 0$ (it corresponds to the
disconnected part of $I_{\alpha\beta}$ described by a diagram 
with no impurity lines at all).
The rest is represented by a sum over closed classical paths 
(Eq.~(\ref{IGG}) and Fig.~\ref{fig3}a) {\it with the direct zero-length 
trajectory excluded}.
This is in full analogy with the semiclassical
consideration of the spectral correlator 
$\langle \nu(E) \nu(E+\omega) \rangle $ \cite{argaman},
where the zero length trajectory yields the disconnected part
$\langle \nu \rangle^2$, while the sum over periodic orbits reproduces 
the Altshuler--Shklovskii diffuson contribution. 
Note that the procedure used by
TS\&E would produce an additional spurious contribution of
very short, almost straight paths in this case as well.

We now turn to the closely related analysis (already presented in a brief
form in the Comment \cite{comment}) of the new term
in the $\sigma$--model obtained by TS\&E in
Refs.~\cite{taras,taras1}. According to these authors,
this term leads to delocalization at the one loop level,
so that the corresponding diagrams should contain
only one true (singular at low momenta and frequency) diffuson.
One can show, however, that, very generally, such a contribution to
conductivity does not diverge. For white-noise RP this
was shown in Ref.~\cite{vw}. We will now prove that the statement
holds also in the case of RMF or smooth RP with time reversal
invariance broken. Indeed, the sum of all one-diffuson
diagrams with $n$ impurity lines crossing the diffuson and having their
starting (end) point anywhere on the left (right) block has the form
\be
\Delta\sigma^{(n)}_{xx}(\omega)=
\frac{e^2}{2\pi} \int (d\bq) \frac{M^{(n)}_{xx}(\bq,\omega)}
{2\pi\nu\tau^2(Dq^2-i\omega)},
\label{cond}
\ee
where
\bea
M_{xx}^{(n)}(\bq,\omega)&=&\frac{1}{n!}\int (d\bq_1)\ldots (d\bq_{n-1})
W(q_1)\ldots W(q_n) 
\nonumber \\ &\times & 
S_{x}^{\rm L}(\bq_1,\ldots,\bq_n) 
S^{\rm R}_{x}(-\bq_1,\ldots,-\bq_n), \ \   {\rm RP}
\nonumber
\\
M_{xx}^{(n)}(\bq,\omega)&=&\frac{1}{n!}\int (d\bq_1)\ldots
(d\bq_{n-1})\nonumber \\ &\times &
{\cal W}_{\alpha_1\beta_1}(\bq_1)\ldots
{\cal W}_{\alpha_n\beta_n}(\bq_n) \nonumber \\
&\times & S^{\rm L}_{x \alpha_1 \ldots \alpha_n}(\bq_1,\ldots,\bq_n)
\nonumber\\ &\times&
S^{\rm R}_{x \beta_1 \ldots \beta_n}(-\bq_1,\ldots,-\bq_n),\ \ 
{\rm RMF}. 
\label{MRMF}
\eea
Here $S^{\rm L}$ ($S^{\rm R}$) denotes the left (right) vertex
part which has one external 
current vertex and to which $n$ impurity lines (momenta $\bq_i$)
as well as one diffuson (momentum $\bq=\sum \bq_i$) are attached.
Using the same arguments as presented above for the vertex part $T$
of the current correlator, one finds \cite{comment}
$S^{\rm L,R}\propto q$ for RP and 
$S^{\rm L,R}\propto q q_1 \ldots q_n$ for RMF, so
that the correction (\ref{cond}) to conductivity is
non-divergent as $\omega \to 0$, independently of the type of disorder
and presence or absence of the time-reversal symmetry, 
\be
\Delta\sigma^{(n)}_{xx}(\omega)\propto
\int (d\bq) \frac{q^2}
{(Dq^2-i\omega)}
\propto {\rm const}+|\omega|\ .
\label{WMPEW}
\ee

In fact, the absence of a divergent correction is intimately
related to the above analysis of the $q\to 0$ form of the current correlator.
Indeed, summing up diagrams with a diffuson vertex inserted in
all possible ways and using
$$
G_E^R G_E^A =i\tau\left[G_E^R-G_E^A \right]
$$
one finds
\be
e^2M_{xx}^{(n)}(q=0,\omega)=(2\pi\tau)^2I_{xx}^{(n)}(q=0,\omega),
\label{MandI}
\ee
so that the prefactor in front of the logarithmic divergency
$\int(d\bq)/\nu(Dq^2-i\omega)$ in (\ref{cond}) is equal to
$I_{xx}^{(n)}(q=0,\omega)$ (as also found by TS\&E).
Since, however, $I_{xx}^{(n)}(q=0,\omega)=0$, this just amounts
to saying that the divergent one-diffuson contribution does not
exist, neither for the RP nor for the RMF problem.
As to the error in the $\sigma$--model calculation of conductivity by
TS\&E in \cite{taras,taras1},
it is identically the same as in their calculation of the current
correlator (discussed above in great detail), in view of the relation
(\ref{MandI}) between the two quantities.

As is well known, divergent contributions to the conductivity
appear if one allows for more than one diffuson. In particular,
the counterpart of the one-- and two--diffuson diagrams for
the current correlator (Fig.~\ref{fig1}) are the conventional two-loop
diagrams for conductivity with two and three diffusons,
yielding the usual weak-localization correction in the unitary
ensemble,
\be
{\Delta g \over g} = -\frac{1}{2\pi^2 g^2}\ln\frac{L_\omega}{l_{\rm tr}},
\label{WL}
\ee
where $g=\sigma_{xx}/(e^2/h)$.

Finally, let us return to the one-diffuson corrections to conductivity
(\ref{WMPEW}). Though non-divergent, they are still of considerable
interest, because of their non-analytic ($\propto |\omega|$) character
at $\omega \to 0$. 
Such contributions correspond to $1/t^2$ long-time tails
in the velocity correlation function and are induced by return processes
neglected in the Boltzmann transport theory. They were discovered
long ago in the context of the classical Lorentz gas
\cite{lorentz-gas} 
and were recently studied \cite{wilke} for a 2D electron gas subject
to a smooth RP or RMF. It was found that, in contrast
to the universal weak localization corrections, the prefactor and the
sign of the $|\omega|$ term is non-universal (i.e. depends on the type
of disorder).

In conclusion, we have studied long-range correlations of local
current densities  ${\bf j}(\br,E)$ in a 2D mesoscopic system with
time reversal invariance broken by a homogeneous or random magnetic
field. We have considered two types of disorder: a (possibly smooth)
random potential and a random magnetic field. We have demonstrated
that within any given order of the perturbation theory  the range of
correlations remains 
finite even in the RMF case, despite the  IR-singular nature of the
vector potential correlator.  The long-range correlations are
determined by the diagrams with diffusons,  Fig.~\ref{fig1}, yielding the
result (\ref{diffres}). The current correlation function is found to
be universal, i.e. it does
not depend on the type or correlation length of disorder.

Our results demonstrate that recent findings of TS\&E
\cite{taras,taras1} 
on the current correlations and on the conductivity renormalization
leading to delocalization in 2D electron systems with a smooth RP or
a RMF (as well as a similar statement of an earlier
paper \cite{zhang}) are fundamentally wrong.
To make this point as clear as possible, we
have demonstrated how our results are reproduced within the ballistic
$\sigma$--model approach. We have further shown that the spurious
contribution found by TS\&E results from an incorrect regularization
procedure violating current conservation and gauge invariance.
This term does not arise if one uses diagrammatic
perturbation theory or a properly regularized
ballistic $\sigma$--model. Therefore, the new term in the diffusive
$\sigma$--model which was obtained in \cite{zhang,taras,taras1}
and which was claimed there to lead to delocalization, does not
exist. In particular, the RMF problem belongs
to the conventional unitary symmetry class
with the leading quantum correction being of two-loop
order and of localizing nature as found earlier \cite{amw,network}.


This work was supported by the SFB195 der Deutschen
Forschungsgemeinschaft, the DFG-Schwerpunktprogram
``Quanten-Hall-Systeme'', the INTAS grant 97-1342,
and the RFBR grant 99-02-17110.

\end{multicols}


\begin{references}
\bibitem{footnote1} The formula (\ref{betaRP}) is not written
explicitly in Refs.~\cite{taras,taras1}, but it follows from Eq.~(54)
of \cite{taras1} if the eigenvalues of the collision
operator characteristic for a RP with large correlation length, 
$w_m \equiv \tau^{-1}(1-\lambda_m) \simeq [(\tau_{\rm tr}/m^2)+\tau]^{-1}$,
are used.
\bibitem{footnote} One should exercise caution
when applying the statement $F(q,\omega)\propto q^2$
at low $q$ to the diffusion contribution, since in this case
the order of limits $q \to 0$ and $\omega \to 0$ is crucial.
If one considers the limit  $q \to 0$ at fixed $\omega$,
then the consideration leading to Eqs.~(\ref{FRPq2}),\ (\ref{FRMFq2})
applies, since $\omega$ effectively sets the maximal length of the
diffusion ladder. In the opposite case, $q\gg L_\omega^{-1}$, one
gets the result (\ref{diffres}). This is, however, irrelevant for
the short-scale ($Q>l_{\rm tr}$) contribution
considered by TS\&E.
\end{references}
\end{document}